\documentclass[11pt,a4paper]{article}
\usepackage{amssymb}
\usepackage[russian,english]{babel}
\usepackage[dvips,unicode]{hyperref}
\begin{document}
\title{Electromagnetic radiation in even-dimensional  
spacetimes}
\author{B. P. Kosyakov}
\maketitle
\begin{center}
{\it Russian Federal Nuclear Center, Sarov, 607190 Nizhnii
Novgorod Region, Russia\\
{\rm E-mail:} ${\rm kosyakov@vniief.ru}$} 
\end{center}
\begin{abstract}
\noindent
The basic concepts and mathematical constructions of the Maxwell--Lorentz 
electrodynamics in flat spacetime of an arbitrary even 
dimension
$d=2n$ are briefly reviewed. 
We show that the retarded field 
strength ${\cal F}^{(2n)}_{\mu\nu}$ 
due to a point charge living in a
$2n$-dimensional 
world
can be algebraically expressed in terms of the 
retarded vector potentials ${\cal A}^{(2m)}_{\mu}$ 
generated by this charge                                         
as if it were accommodated in
$2m$-dimensional 
worlds nearby, $2\le m\le n+1$.
With this finding, the rate of radiated 
energy-momentum of the electromagnetic field takes a compact form. 

\end{abstract}



\section{Introduction}
This paper is dedicated to Professor 
Iosif Buchbinder in celebration of his sixtieth birthday.
A marvellous feature of my friend Iosif is his ability to grasp the 
essence 
of a challenging
problem in theoretical physics and interpret it quite plainly.
He makes a major effort to attain 
the greatest possible clarity 
in a complex subject.
The analysis of the 
Maxwell--Lorentz electrodynamics in even-dimensional 
spacetimes presented in this paper will hopefully be found to be made
in the same vein. 

The physics in higher spacetime dimensions is of basic current interest.
String-inspired large extra-dimensional models \cite{Arkani-Hamedb}, 
\cite{Antoniadis},
\cite{Arkani-Hamedb-prd}
and braneworld scenarios 
\cite
{Akama}
\cite
{Rubakov},
\cite{Visser},
\cite{Randall},
\cite{Randall-s}
 (for a review see
\cite{Rubakov-usp},
\cite{Maartens}) offer promise for a better understanding of a rich
variety of high-energy phenomena which is expected to be discovered 
at the Large Hadron Collider at CERN, and other coming into service colliders.
Our main concern in this paper is with the concept of radiation in 
higher-dimensional classical electrodynamics.
A further refinement of this concept is needed if we are 
to gain a more penetrating insight into the self-interaction problem \cite{k92}, \cite{k06}, \cite{Mironov-Morozov}. 
Recently, this problem was addressed in Refs.~\cite{k99},
\cite{Gal'tsov},
\cite{Kazinski},
\cite{Yaremko}  
\cite{Sarioglu},
\cite{MironovMorozov}.
It should be stressed that spacetime manifolds in these papers were assumed to be
flat. 
(Conceivably it might be prematurely to embark on a study of the radiation in 
curved manifolds 
until the energy-momentum problem in general relativity
is completely solved.) 
We also note that  
the idea of radiation in odd-dimensional 
worlds falls far 
short of being clear-cut because Huygens's principle fails in odd
spacetime dimensions \cite{IwanenkoSokolow},
and the same is true for massive vector fields.
Consequently, we will restrict our consideration to 
classical electrodynamics in Minkowski spacetime of even 
dimension
$d=2n$.

The paper is organized as follows.
Section 2 outlines the state of the art of the $2n$-dimensional
Maxwell--Lorentz theory, notably
the methods for solving 
Maxwell's equations with the source composed of a single point charge.
A central result of this section is given by 
equations (\ref{F-2})--(\ref{F-10})
suggesting that the retarded field 
strength ${\cal F}^{(2n)}_{\mu\nu}$ 
due to a point charge living in a
$2n$-dimensional 
world
can be algebraically expressed in terms of the 
retarded vector potentials ${\cal A}^{(2m)}_{\mu}$ 
generated by this charge                                         
as if it were accommodated in
$2m$-dimensional 
worlds nearby, with $m$ being within the limits $2\le m\le n+1$.
It is then shown in Sec.~3 that
the rate of radiated 
energy-momentum of the electromagnetic field in $2n$-dimensional spacetime
takes a compact form, Eqs.~(\ref{radiat-rate}) and (\ref{rad-rate}). 
Some implications of these results are discussed in Sec.~4. 

We adopt the metric of the form $\eta_{\mu\nu}={\rm diag}\,(1,-1,\ldots,-1)$, and follow the conventions of 
Ref. \cite{k06} throughout.

\section{Vector potentials, prepotentials, and field strengths}
\label{ret}
Consider a single charged point particle moving along a timelike 
world line in flat spacetime of an arbitrary even dimension $d=2n$, 
$n=1,2,\ldots$ 
With reference to the aforementioned 
string-inspired models and 
braneworld scenarios,
our prime interest is with $d$ in the range from $d=2$ to $d=10$.
The world line $z^\mu(s)$  is regarded as a smooth function of the 
proper time $s$.
Suppose that the 
Maxwell--Lorentz electrodynamics is still valid.
This is tantamount to stating that
the field sector 
is given by 
\begin{equation}
{\cal L}=
-
{1\over 4\Omega_{d-2}}\,F_{\mu\nu}F^{\mu\nu}- A_\mu j^\mu,
\label
{Lagrangian-Maxw-D}
\end{equation}                                          
\begin{equation}
j^\mu(x)=e\int_{-\infty}^\infty ds\,v^\mu(s)\,\delta^{\hskip0.2mm d}
\left[x-z(s)\right],
\label
{single-charge-j}
\end{equation}                        
and the retarded boundary 
condition is imposed on the vector potential $ A_\mu$.
Here, 
$\Omega_{d-2}$ is                                          
the area of the unit $(d-2)$-sphere, 
$v^\mu={\dot z}^\mu=dz^\mu/ds$ 
is the $d$-velocity, and $\delta^{\hskip0.2mm d}(R)$ is
the $d$-dimensional Dirac delta-function.
In what follows the value of the charge will be taken to be unit, $e=1$.

The field equation resulting from (\ref{Lagrangian-Maxw-D}) reads 
\begin{equation}
{\cal E}^\mu
=
\partial_\nu F^{\mu\nu}
+
\Omega_{d-2} j^\mu
=0.
\label
{maxw-D}
\end{equation}                                          
This is accompanied by the Bianchi identity 
\begin{equation}
{\cal E}^{\lambda\mu\nu}
=\partial^\lambda F^{\mu\nu}+\partial^\nu F^{\lambda\mu}+
\partial^\mu F^{\nu\lambda}=0.
\label
{maxw}
\end{equation}                                            

We take the general solution to (\ref{maxw}),
$F_{\mu\nu}=\partial_\mu A_\nu-\partial_\nu A_\mu$, and choose the 
Lorenz gauge condition $\partial_\mu A^\mu=0$ to put
(\ref{maxw-D}) into the form
\begin{equation}
\Box\, A^\mu=\Omega_{d-2}\,j^\mu.
\label
{wave-equation-D}
\end{equation}                                          

There are two alternative procedures for integrating the wave equation
(\ref{wave-equation-D}). 
The Green's function approach  
holds much favor (for a review see \cite{IwanenkoSokolow},  
\cite{k06}).
The retarded Green's function
satisfying  
\begin{equation}
\Box\, G_{\rm ret}(x)=\delta^{d}(x)
\label
{wave-equation-Green-D}
\end{equation}                                          
is given by
\begin{equation}
G_{\rm ret}(x)
={
\frac{1}{2\pi^n}\, 
\theta(x_0)\,\delta^{(n-1)}(x^2), 
\quad
 d=2n. 
}
\label
{Green's-D}
\end{equation}
Here, $\delta^{(n-1)}(x^2)$ 
is  
the delta-function
differentiated $n-1$ times with respect to its
argument.
With the Green's function (\ref{Green's-D}) at our disposal it 
becomes possible to obtain the retarded vector potential
\begin{equation}
A^\mu(x)
=\Omega_{d-2}\int_{-\infty}^{\infty} ds\,G_{\rm ret}\left(R\right)v^\mu(s), 
\label
{vector-potential-Green's-D}
\end{equation}
where 
\begin{equation}
R^\mu=x^\mu-z^\mu(s) 
\label
{R-mu-df}
\end{equation}
 is the null four-vector 
drawn from the retarded point $z^\mu(s)$ on the world line, where the signal is emitted,
to the point $x^\mu$, where the signal is received.

The other procedure consists of 
using the ansatz of a particular form \cite{k99}, \cite{k06}.
To illustrate,  the pertinent ans{\"a}tze  for $d=2,4,6$
are given, 
respectively, by
\begin{equation}
A^{(2)}_\mu=\alpha(\rho)\,R_\mu,
\label
{2-Ansatz}
\end{equation}                                     
\begin{equation}
A^{(4)}_\mu=f(\rho)\,R_\mu +
g(\rho)\,v_\mu,
\label
{4-Ansatz}
\end{equation}                                     
\begin{equation}
A^{(4)}_\mu
=\Omega(\rho,\lambda)\,R_\mu+\Phi(\rho,\lambda)\,v_\mu+
\Psi (\rho,\lambda)\, a_\mu.
\label
{6-Ansatz}
\end{equation}                                     
Here,
 $a_\mu={\dot v}_\mu$ is the $d$-acceleration, and $\alpha$, $f$, $g$, 
$\Omega$, $\Phi$, $\Psi$ are unknown scalar functions. 
The functions $\alpha$, $f$, $g$ are assumed to depend on the retarded
invariant distance
\begin{equation}
\rho=R\cdot v,
\label
{rho-def}
\end{equation}                                          
while $\Omega$, $\Phi$, $\Psi$ are taken to depend on $\rho$ and the retarded
invariant variable 
\begin{equation}
\lambda=R\cdot a-1.
\label
{lambda-def}
\end{equation}                                          
Let us introduce a further null vector $c_\mu$ aligned with $R_\mu$,
\begin{equation}
R_\mu
={\rho}\,{c_\mu}.
\label
{c-mu-def}
\end{equation} 
We insert any one of the ans{\"a}tze (\ref{2-Ansatz}),  (\ref{4-Ansatz}), 
(\ref{6-Ansatz}) in (\ref{maxw-D}),  perform
differentiations  of the retarded variables 
using  the rules 
\begin{equation}
\partial_\mu s=c_\mu,
\label
{d-tau}
\end{equation}                                          
\begin{equation}
\partial_\mu\rho=v_\mu+ \lambda c_\mu,
\label
{partial-mu-rho=}
\end{equation}                                          
\begin{equation}
\partial_\mu R^\lambda= 
\delta^\lambda_{~\mu}-v^\lambda\,c_\mu ,
\label
{X-diff}
\end{equation}                                                                                   
and solve the resulting ordinary differential equations 
(for detail see \cite{k99}, \cite{k06}) to obtain
\begin{equation}
{A}^{(2)}_\mu=-R_\mu,
\label
{A-2}
\end{equation}                       
\begin{equation}
F^{(2)}_{\mu\nu}=c_\mu v_\nu-c_\nu v_\mu, 
\label
{F-LW-2}
\end{equation}                       
\begin{equation}
{A}^{(4)}_\mu=\frac{v_\mu}{\rho}\,,
\label
{A-4}
\end{equation}                       
\begin{equation}
F^{(4)}_{\mu\nu}=c_\mu U^{(4)}_\nu-c_\nu U^{(4)}_\mu, 
\quad
U^{(4)}_\mu=-\lambda\,\frac{v_\mu}{\rho^2} +\frac{a_\mu}{\rho}\,,
\label
{U-4}
\end{equation}                       
\begin{equation}
{A}^{(6)}_\mu=\frac13\left(-\lambda\,\frac{v_\mu}{\rho^3} 
+\frac{a_\mu}{\rho^2}\right)\,,
\label
{A-6}
\end{equation}                       
\begin{equation}
F^{(6)}_{\mu\nu}=
\frac13\left(c_\mu U^{(6)}_\nu-
c_\nu U^{(6)}_\mu+\frac{a_\mu v_\nu-a_\nu v_\mu}{\rho^3}\right),
\quad
U^{(6)}_\mu 
=\left[3\lambda^2-\rho^2\,({\dot a}\cdot c)\right]\frac{v_\mu}{\rho^4}
-3\lambda\,\frac{a_\mu}{\rho^3}
+\frac{{\dot a}_\mu}{\rho^2}.
\label
{6-vector-V}
\end{equation}                                           

Note the overall factor $\frac13$ in (\ref{A-6}).
The origin of this numerical 
factor is most easily understood if we apply Gauss'
law to the case that $a_\mu=0$ and
${\dot a}_\mu=0$.
To simplify our notations as much as possible, we introduce the
{\it net} vector potentials and field strengths, ${\cal A}_{\mu}$ and 
${\cal F}_{\mu\nu}$ (as opposed to the ordinary vector potentials 
and field strengths, ${A}_{\mu}$ and ${F}_{\mu\nu}$, whose {normalization} is consistent 
with Gauss' law):
\begin{equation}
{A}^{(2p)}_\mu
=N_p^{-1}{\cal A}^{(2p)}_\mu,
\quad
{F}^{(2p)}_{\mu\nu}
=N_p^{-1}{\cal F}^{(2p)}_{\mu\nu},
\label
{A-cal-A}
\end{equation}                       
where
\begin{equation}
N_p=(p-1)!!\,.
\label
{N-p-df}
\end{equation}                       

It is an easy matter to extend the sequence of the ans{\"a}tze shown 
in (\ref{2-Ansatz}),  
(\ref{4-Ansatz}), (\ref{6-Ansatz}) to any $d=2n$ with integer $n\ge 1$.
Based on the anzatz for $d=2n$, we come to the anzatz for $d=2n+2$ by
appending a term proportional 
to the $(n-1)$th derivative of $v_\mu$ with respect to $s$, and assuming 
that the unknown functions
depend on $\rho$, together with scalar products of $R_\mu$ and derivatives of
$v_\mu$ up to the $(n-1)$th derivative inclusive. 

Proceeding in these lines, we get 
\begin{equation}
{\cal A}^{(2)}_\mu=-R_\mu,
\label
{cal-A-2}
\end{equation}                       
\begin{equation}
{\cal A}^{(4)}_\mu=\frac{v_\mu}{\rho}\,,
\label
{cal-A-4}
\end{equation}                       
\begin{equation}
{\cal A}^{(6)}_\mu=-\lambda\,\frac{v_\mu}{\rho^3} +\frac{a_\mu}{\rho^2}\,,
\label
{cal-A-6}
\end{equation}                       
\begin{equation}
{\cal A}^{(8)}_\mu
=\left[3\lambda^2
-\rho^2\left({\dot a}\cdot c\right)\right]\frac{v_\mu}{\rho^5} 
-3\lambda\,\frac{a_\mu}{\rho^4}
+\frac{{\dot a}_\mu}{\rho^3}\,,
\label
{cal-A-8}
\end{equation}                       
\begin{equation}
{\cal A}^{(10)}_\mu
=\left[-15\lambda^3
+10\lambda\rho^2\left({\dot a}\cdot c\right)-\rho^2a^2
-\rho^3\left({\ddot a}\cdot c\right)\right]\frac{v_\mu}{\rho^7} 
+\left[15\lambda^2
-4\rho^2\left({\dot a}\cdot c\right)\right]\frac{a_\mu}{\rho^6}
-6\lambda\,\frac{{\dot a}_\mu}{\rho^5}
+\frac{\ddot a_\mu}{\rho^4}\,,
\label
{cal-A-10}
\end{equation}                       
$$
{\cal A}^{(12)}_\mu
=\left\{105\lambda^2\left[\lambda^2
-\rho^2({\dot a}\cdot c)\right]
+ 15\lambda\rho^2[\rho({\ddot a}\cdot c)+a^2]
-\frac52\,\rho^3({a}^2)^{.}
-\rho^4(\stackrel{\ldots}{a}\cdot c)
+10\rho^4({\dot a}\cdot c)^2
\right\}\frac{v_\mu}{\rho^9} 
$$
\begin{equation}
+5\left\{3\lambda\left[-7\lambda^2
+4\rho^2({\dot a}\cdot c)\right]
-\rho^2[\rho({\ddot a}\cdot c)
+ a^2]\right\}\frac{{a}_\mu}{\rho^8}
+5\left[9\lambda^2
-2\rho^2({\dot a}\cdot c)
\right]\frac{{\dot a}_\mu}{\rho^7}
-10\lambda\,\frac{{\ddot a}_\mu}{\rho^6}
+\frac{{\stackrel{\ldots}{a}}_\mu}{\rho^5}\,.
\label
{cal-A-12}
\end{equation}                       

Another way of looking at ${\cal A}^{(2p)}_\mu$ is to
invoke the notion of prepotential.
The prepotential ${H_\mu}$ of the vector potential $A_\mu$ is
defined as
\begin{equation}
{A}_\mu
=
\Box\,
{H}_\mu.
\label
{H-mu-def}
\end{equation}                       
One can check that 
\begin{equation}
N_{p+1}\,\Box\,
{A}^{(2p)}_\mu
=
{\left(d-2p\right)N_{p}}\,
{A}^{(2p+2)}_\mu,
\quad
p\ge 1.
\label
{A-2p-A-2p-1}
\end{equation}                       
In other words, any $2n$-dimensional retarded vector potential $A^{(2n)}_\mu$  
(up to
a normalization factor) is the prepotential of 
the $(2n+2)$-dimensional retarded vector potential $A^{(2n+2)}_\mu$. 
Furthermore, ${\cal A}^{(2p)}_\mu$ can be produced by acting on ${\cal A}^{(2)}_\mu$
$p-1$ times with the wave operator:
\begin{equation}
{\cal A}^{(2p)}_\mu
=
Z^{-1}_{d,p}\,
\Box^{p-1}
{R}_\mu
=
-Z^{-1}_{d,p}\,\Box^{p-1}
{\cal A}^{(2)}_\mu,
\label
{A-2p-A-2}
\end{equation}                       
where
\begin{equation}
Z_{d,p}
=
\left(d-2\right)
\left(d-4\right)\cdots
\left(d-2p\right)
N_p=\frac{2^p\left(n-1\right)!}{(n-p-1)!}\,N_p.
\label
{Z-def}
\end{equation}                       
All the resulting vector potentials ${\cal A}^{(2p)}_\mu$, beginning with $p=2$,
obey the Lorenz gauge condition.
To see this, we note that $\partial^\mu R_\mu=d-1$, and so $\Box\,
\partial^\mu {\cal A}^{(2)}_\mu=0$.

This technique provides a further significant advantage if we observe that
the action of the wave operator amounts to the action of 
the first-order differential operator
\begin{equation}
\frac{1}{\rho}\,
\frac{d}{ds}\,.
\label
{A-2p+2-A-2p-diff}
\end{equation}                       
We thus have
\begin{equation}
{\cal A}^{(2p)}_\mu
=
-\left(\frac{1}{\rho}\,
\frac{d}{ds}\right)^{p-1}
{\cal A}^{(2)}_\mu.
\label
{A-2p+2-A-2-diff}
\end{equation}                       
Indeed, (\ref{A-2p+2-A-2-diff}) derives from (\ref{Green's-D}) and 
(\ref{vector-potential-Green's-D}) by noting that
$dR^2/ds=-2\rho$,
$dR_\mu/ds=-v_\mu$, and so
\begin{equation}
-\frac{1}{\rho}\,
\frac{d}{ds}\,
{\cal A}^{(2)}_\mu
=
\frac{1}{\rho}\,
\frac{d}{ds}\,
{R}_\mu
=
\frac{v_\mu}{\rho}
={\cal A}^{(4)}_\mu.
\label
{A-4-A-2-diff}
\end{equation}                       

We now take a closer look at the field strengths $F^{(2)}_{\mu\nu}$,
$F^{(4)}_{\mu\nu}$, and $F^{(6)}_{\mu\nu}$ shown, respectively, in
(\ref{F-LW-2}), (\ref{U-4}), and (\ref{6-vector-V}).
When their structure is compared with that of the vector potentials 
${\cal A}^{(2)}_{\mu}$, ${\cal A}^{(4)}_{\mu}$, ${\cal A}^{(6)}_{\mu}$, 
${\cal A}^{(8)}_{\mu}$
displayed in (\ref{cal-A-2})--(\ref{cal-A-8}), it is apparent that
\begin{equation}
{\cal F}^{(2)}
=-{\cal A}^{(2)}\wedge{\cal A}^{(4)} ,
\label
{F-2}
\end{equation}                       
\begin{equation}
{\cal F}^{(4)}
=-{\cal A}^{(2)}\wedge{\cal A}^{(6)} ,
\label
{F-4}
\end{equation}                       
\begin{equation}
{\cal F}^{(6)}
=
-{\cal A}^{(2)}\wedge{\cal A}^{(8)} 
-
{\cal A}^{(4)}\wedge{\cal A}^{(6)}.
\label
{F-6}
\end{equation}                       
In addition, one can verify that
\begin{equation}
{\cal F}^{(8)}
=
-{\cal A}^{(2)}\wedge{\cal A}^{(10)} 
-
2{\cal A}^{(4)}\wedge{\cal A}^{(8)} ,
\label
{F-8}
\end{equation}                       
\begin{equation}
{\cal F}^{(10)}
=
-{\cal A}^{(2)}\wedge{\cal A}^{(12)} 
-
3{\cal A}^{(4)}\wedge{\cal A}^{(10)}
-
2{\cal A}^{(6)}\wedge{\cal A}^{(8)}. 
\label
{F-10}
\end{equation}                       
We come to recognize that the retarded field strength 
${\cal F}^{(2p)}_{\mu\nu}$ can be expressed 
in a very compact and elegant form
in terms of  
retarded vector potentials 
${\cal A}^{(2m)}_{\mu}$,  $2\le m\le p+1$.
Recall, the canonical representation of a general 2-form $\omega^{(2n)}$ in 
spacetime of
dimension $d=2n$ is the sum of $n$ exterior products of 1-forms:
\begin{equation}
{\omega^{(2n)}}
=
{f}_1\wedge{f}_{2} 
+
\cdots
+
{f}_{2n-1}\wedge{f}_{2n}. 
\label
{omega-canonical-decomp}
\end{equation}                       
In particular, by (\ref{omega-canonical-decomp}), $\omega^{(10)}$ is decomposed into the sum involving 
five terms.
However, (\ref{F-10}) shows that the retarded field
strength contains only three exterior products, two less than 
the canonical representation.

The validity of relations (\ref{F-2})--(\ref{F-10}) can be seen
by inspection.
To derive them in a regular way, we take (\ref{F-2}) as the 
starting point.
If we apply $Z^{-1}_{d,p}\,\Box^{p-1}$ to the left-hand side of this equation,
then, in view of
(\ref{A-2p-A-2}), we obtain ${\cal F}^{(2p)}$.
Applying $p-1$ times the first-order differential operator
(\ref{A-2p+2-A-2p-diff}) to the right-hand side of (\ref{F-2}) and
taking into account Leibnitz's rule 
for differentiation of the product of
two functions,
in view of (\ref{A-2p+2-A-2-diff}),
we come to the desired result.

This explains the puzzling fact that the
gauge-independent quantity ${\cal F}^{(2p)}$ is an algebraic function of
gauge-dependent quantities ${\cal A}^{(2m)}$.
By the construction, the vector potentials 
${\cal A}^{(2m)}_\mu$, $m\ge 1$, are
subject to the 
Lorenz gauge condition.
Therefore, such
${\cal A}^{(2m)}_\mu$ leave room for gauge modes $\partial_\mu\chi$ 
with $\chi$ being solutions to the wave equation, $\Box\chi=0$.
In our derivation of (\ref{F-4})--(\ref{F-10}), we are 
entitled to apply the wave operator $\Box$,  rather than the  
first-order differential operator 
(\ref{A-2p+2-A-2p-diff}), to the right-hand 
side of 
(\ref{F-2}).
All feasible gauge modes are then killed by the action of $\Box$.

We close this section with a remark about the behavior of the
retarded electromagnetic field 
at spatial infinity.
In general, ${\cal F}^{(2n)}$ can be represented as the sum of  
exterior products of retarded vector potentials 
${\cal A}^{(2p)}\wedge{\cal A}^{(2n-2p+4)}$.
It is easy to understand that 
the infrared properties of ${\cal F}^{(2n)}$ are controlled by the term
${\cal A}^{(2)}\wedge{\cal A}^{(2n+2)}$. 
(In fact, a comparison of the long-distance behavior of 
${\cal A}^{(2)}\wedge{\cal A}^{(2n+2)}$ and 
${\cal A}^{(4)}\wedge{\cal A}^{(2n)}$ 
will suffice 
for the present purposes.
Since the least falling terms of ${\cal A}^{(2n+2)}$ and 
${\cal A}^{(2n)}$ scale, respectively, as $\rho^{-n}$ and $\rho^{1-n}$, 
the leading long-distance asymptotics of 
${\cal A}^{(2)}\wedge{\cal A}^{(2n+2)}$ is given
by $\rho^{1-n}$ while that of
${\cal A}^{(4)}\wedge{\cal A}^{(2n)}$ is given by 
$\rho^{-n}$.) 
We segregate in ${\cal A}^{(2n+2)}$ the term scaling as
$\rho^{-n}$ by introducing the vectors  
\begin{equation}
{\mathfrak b}^{(2n+2)}_\mu
=
\lim_{\rho\to\infty}\,\rho^{n}{\cal A}^{(2n+2)}_\mu
\label
{infrare-a}
\end{equation}                       
and
\begin{equation}
{\bar{\cal A}}^{(2n+2)}_\mu
=
\frac{1}{\rho^n}\,{\mathfrak b}^{(2n+2)}_\mu.
\label
{infrared-A}
\end{equation}                       
All infrared irrelevant terms are erased by
this limiting procedure,  so that
\begin{equation}
{{\cal A}}^{(2)}
\wedge 
{\bar{\cal A}}^{(2n+2)}
\label
{infrared-F}
\end{equation}                       
represents the infrared part of ${\cal F}^{(2n)}$.

We write explicitly ${\mathfrak b}^{(2n+2)}_\mu$ for $n=1,2,3,4,5$: 
\begin{equation}
{\mathfrak b}^{(4)}_\mu
=v_\mu,
\label
{mf-4}
\end{equation}                       
\begin{equation}
{\mathfrak b}^{(6)}_\mu
=-
\left({a}\cdot c\right)v_\mu
+
{a_\mu},
\label
{mf-6}
\end{equation}                       
\begin{equation}
{\mathfrak b}^{(8)}_\mu
=
\left[3\left({a}\cdot c\right)^2
-\left({\dot a}\cdot c\right)
\right]{v_\mu} 
-3\left({a}\cdot c\right){a_\mu}
+{{\dot a}_\mu},
\label
{mfr-8}
\end{equation}                       
\begin{eqnarray}
{\mathfrak b}^{(10)}_\mu
=-\left[15\left({a}\cdot c\right)^3
-10\left({a}\cdot c\right)\left({\dot a}\cdot c\right)
+\left({\ddot a}\cdot c\right)\right]{v_\mu} 
\nonumber\\
+\left[15\left({a}\cdot c\right)^2
-4\left({\dot a}\cdot c\right)\right]{a_\mu}
-6\left({a}\cdot c\right){{\dot a}_\mu}
+{\ddot a_\mu},
\label
{mfr-10}
\end{eqnarray}                       
\begin{eqnarray}
{\mathfrak b}^{(12)}_\mu
=\left\{
5\left[
3\cdot 7\left({a}\cdot c\right)^2\left(\left({a}\cdot c\right)^2
-\left({\dot a}\cdot c\right)\right)
+2\left({\dot a}\cdot c\right)^2
+ 
3\left({a}\cdot c\right)\left({\ddot a}\cdot c\right)
\right]
-(\stackrel{\ldots}{a}\cdot c)
\right\}
{v_\mu} 
\nonumber\\
-5\left\{3\left({a}\cdot c\right)
[7\left({a}\cdot c\right)^2
-
4({\dot a}\cdot c)]
+({\ddot a}\cdot c)\right\}{{a}_\mu}
+5\left[9\left({a}\cdot c\right)^2
-2({\dot a}\cdot c)\right]{\dot a}_\mu
-10\left({a}\cdot c\right){\ddot a}_\mu
+{\stackrel{\ldots}{a}}_\mu.
\label
{mfr-12}
\end{eqnarray}                       

It follows from (\ref{mf-6})--(\ref{mfr-12}) that 
${\mathfrak b}^{(6)},\ldots,{\mathfrak b}^{(12)}$
are subject to the constraint
\begin{equation}
R\cdot{\mathfrak b}^{(2n+2)}=0,
\label
{constr}
\end{equation}                       
while ${\mathfrak b}^{(4)}$ is not.
To derive (\ref{constr}), 
we 
note that, far apart from the the world 
line, the field appears (locally) as a plane wave moving along a
null ray that points toward the propagation vector $k_\mu$,  
\begin{equation}
{\cal A}_\mu\sim
\epsilon_\mu{\phi}(k\cdot x),
\label
{plane-wave-A}
\end{equation}                       
\begin{equation}
{\cal F}_{\mu\nu}\sim
\left(k_\mu\epsilon_\nu
-k_\nu\epsilon_\mu\right){\phi}'. 
\label
{plane-wave-F}
\end{equation}                       
Here, $\epsilon_\mu$ is the polarization vector, ${\phi}$ is an
arbitrary smooth function of the phase $k\cdot x$, and the prime stands for 
the derivative with respect to the phase.
Recall that   
$\partial^\mu{\cal A}_\mu^{(2n)}=0$ for $n\ge 2$.
In view of (\ref{plane-wave-A}), this equation becomes 
\begin{equation}
\left(k\cdot\epsilon\right){\phi}'=0, 
\label
{plane-wave-Lorenz}
\end{equation}                       
which implies that the polarization vector is orthogonal to the 
propagation vector.
On the other hand, ${\cal A}^{(2n+2)}_\mu$ 
approaches ${\bar{\cal A}^{(2n+2)}_\mu}$ as
$\rho\to\infty$.
Now the null vector $R_\mu$ acts as the propagation vector $k_\mu$. 
A comparison between (\ref{infrared-F}) and (\ref{plane-wave-F}) 
shows that  $\epsilon_\mu$
should be identified with ${\mathfrak b}^{(2n+2)}_\mu$.

Since $\partial^\mu R_\mu=d-1$, the vector potential 
${\cal A}_\mu^{(2)}$ 
does not obey the Lorenz gauge condition, and hence
(\ref{constr}) is not the case for 
${\mathfrak b}^{(4)}_\mu$.

To sum up, the polarization of the retarded
electromagnetic field is an imprint of the next even dimension $d+2$, 
excluding $d=2$ which is immune from the effect 
of $d=4$.

\section{Radiation}
\label{rad}
Apart from the overall numerical
factor,
the {metric stress-energy tensor} 
of the electromagnetic field takes the same form in any dimension,
\begin{equation}
{\Theta}_{\mu\nu}=
\frac{1}{\Omega_{d-2}}\left(F_{\mu}^{~\alpha} 
F_{\alpha\nu}
+\frac{\eta_{\mu\nu}}{4}\,F^{\alpha\beta}F_{\alpha\beta}\right).
\label
{Hilbert-energy-tensor-em}
\end{equation}                         

Let us substitute (\ref{vector-potential-Green's-D}) into 
(\ref{Hilbert-energy-tensor-em}).
Since the result is to be integrated over $(d-1)$-dimensional spacelike 
surfaces, 
${\Theta}^{\mu\nu}$ is conveniently split into two parts, nonintegrable
and integrable,
\begin{equation}
{\Theta}^{\mu\nu}
=
{\Theta}_{\rm I}^{\mu\nu}
+
{\Theta}_{\rm II}^{\mu\nu}.
\label
{energy-tensor-splitting}
\end{equation}                         
Here, our concern is only with the integrable part
${\Theta}_{\rm II}^{\mu\nu}$.
To identify this part of the  stress-energy tensor as the {\it radiation}, 
we check the 
fulfilment of the following conditions \cite{Teitelboim}, \cite{k06}: 

(i) $\Theta^{\mu\nu}_{\rm\hskip0.3mm I}$ and 
$\Theta^{\mu\nu}_{\rm\hskip0.3mm II}$
are dynamically independent off the world line, that is, 
\begin{equation}
\partial_\mu\Theta^{\mu\nu}_{\rm\hskip0.3mm I}=0,
\quad
\partial_\mu \Theta_{{\rm\hskip0.3mm II}}^{\mu\nu}=0,
\label
{cons-Theta-I-II}
\end{equation}                                       

(ii)  $\Theta^{\mu\nu}_{\rm\hskip0.3mm II}$
propagates along the future light cone $C_+$ drawn from the emission 
point,
and 

(iii)  the energy-momentum flux of $\Theta^{\mu\nu}_{{\rm\hskip0.3mm II}}$ 
goes as
$\rho^{2-d}$ implying that the same amount of 
energy-momentum flows 
through spheres of different radii.

It has been found in the previous section that the infrared
behavior\footnote{The term 
`infrared' is used here in reference to what can be described by 
means of  quantities which are either regular or having integrable singularities 
at the world line.} of ${\cal F}^{(2n)}$ is controlled by 
\begin{equation}
{{\cal A}}^{(2)}
\wedge 
{{\cal A}}^{(2n+2)}.
\label
{A-2-wedge-A-2n+2}
\end{equation}                       
More precisely, 
the leading long-distance term 
\begin{equation}
{{\cal A}}^{(2)}\wedge{\bar{\cal A}}^{(2n+2)},
\label
{infrared-F-}
\end{equation}                       
where ${\bar{\cal A}}^{(2n+2)}_\mu$ is defined in (\ref{infrared-A}),
is responsible for the infrared properties of ${\cal F}^{(2n)}$.

With (\ref{Hilbert-energy-tensor-em}), it is apparent that 
${\Theta}_{\rm II}^{\mu\nu}$
is built up solely from the term shown in
(\ref{infrared-F-}),
\begin{equation}
{\Theta}_{\rm II}^{\mu\nu}
=
\frac{-1}{N_n^2\Omega_{2n-2}}\,
R^\mu R^\nu\left({\bar{\cal A}}^{(2n+2)}\right)^2
=
\frac{-1}{N_n^2\Omega_{2n-2}\rho^{2n-2}}\,
c^\mu c^\nu\left({\mathfrak b}^{(2n+2)}\right)^2.
\label
{energy-tensr-rad}
\end{equation}                         
Let us check that ${\Theta}_{\rm II}^{\mu\nu}$ given by 
(\ref{energy-tensr-rad}) meets every condition
(i)--(iii), and hence this quantity
is reasonable to call the radiation\footnote{Strictly speaking, the radiation is represented by 
(\ref{energy-tensr-rad}) only when  $n\ge 2$.
In a world with one temporal and one 
spatial dimension, the radiation is absent \cite{k99}, \cite{k06}.}.
In view of (\ref{infrared-A}), 
the scaling
properties of this ${\Theta}_{\rm II}^{\mu\nu}$ are in agreement with (iii).
Furthermore, since the surface element of the future light cone
${C}_{+}$ is  
\begin{equation}
d\sigma^\mu=c^\mu\rho^{2n-2} d\rho\,d\Omega_{2n-2},
\label
{lc-measure}
\end{equation}                         
where $c^\mu$ is a null vector on $C_+$, the flux of 
$\Theta^{\mu\nu}_{{\rm\hskip0.3mm II}}$ through $C_+$ vanishes, 
$d\sigma_\mu\Theta^{\mu\nu}_{{\rm\hskip0.3mm II}}=0$.
Therefore, $\Theta^{\mu\nu}_{\rm\hskip0.3mm II}$
propagates along $C_+$ to suit (ii).

To verify that condition (i) holds, let us note that, for
$\Theta^{\mu\nu}$ and $j^\mu$ written, respectively, as
(\ref{Hilbert-energy-tensor-em})
and (\ref{single-charge-j}), 
\begin{equation}
\partial_\nu\Theta^{\mu\nu}=- 
 F^{\mu\nu}j_{\nu}.
\label
{div-energy-tensor-nonlin}
\end{equation}                          
Off the world line, (\ref{div-energy-tensor-nonlin}) becomes
\begin{equation}
\partial_\mu \Theta^{\mu\nu}=0,
\label
{cons-Theta}
\end{equation}                                       
and hence either of two local conservation 
laws (\ref{cons-Theta-I-II}) 
implies the other one.
It is sufficient to verify the conservation law for the
$\Theta^{\mu\nu}_{{\rm\hskip0.3mm II}}$.
We have 
\begin{equation}
\partial_\mu {\Theta}_{\rm II}^{\mu\nu}
\propto
\partial_\mu\left[R^\mu R^\nu\left({\bar{\cal A}}^{(2n+2)}\right)^2\right]
= R^\nu\left[2n\left({\bar{\cal A}}^{(2n+2)}
\right)^2
+
\left(R\cdot\partial\right)\left({\bar{\cal A}}^{(2n+2)}\right)^2\right].
\label
{energy-tensor-rad-consrv}
\end{equation}                         
Here the second equation is obtained using the differentiation rule
(\ref{X-diff}) and the fact that
$\delta_{~\mu}^\mu=2n$.
Let us take into
account 
that ${\mathfrak b}^{(2n+2)}_\mu$ depends on  
$v^\alpha$, $a^\alpha$, ... and their scalar products with $c^\alpha$. 
Since
\begin{equation}
(R\cdot\partial)\,\{c^\nu,\,\,v^\nu,\,\,a^\nu,\,\,
{\dot a}^\nu,\ldots\}=0,
\quad
(R\cdot\partial)\,\rho=\rho,
\label
{Prbl4.5.1-10}
\end{equation}                                          
we apply $\left(R\cdot\partial\right)$ to ${\bar{\cal A}}^{(2n+2)}_\mu$ 
defined in (\ref{infrared-A})
to conclude from (\ref{energy-tensor-rad-consrv}) that 
$\partial_\mu {\Theta}_{\rm II}^{\mu\nu}=0$.
This is just the
required result.

By (\ref{constr}),  ${\mathfrak b}^{(2n+2)}_\mu$ is orthogonal to a null
vector $R_\mu$.
This suggests that ${\mathfrak b}^{(2n+2)}_\mu$ 
is a linear combination of a spacelike
vector and the null vector  $R_\mu$ itself.    
Referring to (\ref{infrared-F}),  
${\bar{\cal A}}^{(2n+2)}_\mu$  is defined
up to adding $k\,R_\mu$, where $k$ is an arbitrary constant.
If we impose the Lorenz gauge condition
to this additional term, then 
$k\left(n-1\right)=0$, that is, $k=0$ for $n\ne 1$.
When it is considered that ${\mathfrak b}^{(2n+2)}_\mu$ is spacelike,  
(\ref{energy-tensr-rad}) shows that ${\Theta}_{\rm II}^{00}\ge 0$.  
We thus see that ${\Theta}_{\rm II}^{00}$ represents positive field energy 
flowing
outward from the source.

Let us calculate the radiation rate.
The radiation flux through a $(d-2)$-dimensional sphere enclosing
the source is constant for any radius of the sphere.
Therefore, the terms of $\Theta_{\mu\nu}$ responsible for this flux 
scale as
$\rho^{2-d}$.
The radiated energy-momentum is defined by
\begin{equation}
{\cal P}^{\mu}=
\int_{\Sigma} d\sigma_\nu\,\Theta^{\mu\nu}_{\rm II},
\label
{six-radiat}
\end{equation}                                       
where $\Sigma$ is a $(d-1)$-dimensional spacelike hypersurface.
Since $\Theta^{\mu\nu}_{\rm II}$ involves only integrable singularities, and
$\partial_\nu\Theta^{\mu\nu}_{\rm II}=0$, the surface of integration $\Sigma$ in
(\ref{six-radiat}) may be chosen arbitrarily.
It is convenient to deform $\Sigma$ to a tubular surface 
${T}_\epsilon$ of small invariant radius $\rho=\epsilon$
 enclosing the world line. 
The surface element on this tube is 
\begin{equation}
d\sigma^\mu=\partial^\mu\!\rho\,\rho^{d-2}\,d\Omega_{d-2}\,ds
=(v^\mu+
\lambda c^\mu)\,\epsilon^{d-2}\,d\Omega_{d-2}\,ds.
\label
{tube-measure}
\end{equation}                                       
Equation (\ref{six-radiat})
becomes 
\begin{equation}
{{\cal P}}_{\mu}^{(2n)}=
-\frac{1}{N_n^2\Omega_{2n-2}}
\int^s_{-\infty}ds
\int d\Omega_{2n-2}\, c_\mu 
\left({\mathfrak b}^{(2n+2)}\right)^2,
\label
{radiation-rate}
\end{equation}                         
so that the radiation rate is given by
\begin{equation}
{\dot{\cal P}}_{\mu}^{(2n)}
=
-\frac{1}{N_n^2\Omega_{2n-2}}
\int d\Omega_{2n-2}\, c_\mu 
\left({\mathfrak b}^{(2n+2)}\right)^2.
\label
{radiat-rate}
\end{equation}                         

This can be recast as
\begin{equation}
{\dot{\cal P}}_{\mu}^{(2n)}
=
-\frac{1}{Z^{2}_{d,n}\Omega_{2n-2}}
\int d\Omega_{2n-2}\, c_\mu \left(\lim_{\rho\to\infty}\rho^{n}
\Box^{n}
R_\alpha\right)^2.
\label
{rad-rate}
\end{equation}                         
Conceivably this form of  the radiation rate 
might find use in a wider context of gauge theories.

The solid angle integration is greatly simplified if we introduce the spacelike 
normalized vector $u^\mu$ orthogonal to $v^\mu$, 
\begin{equation}
c^\mu
=
v^\mu
+
u^\mu\,,
\label
{c=v+u}
\end{equation}                         
and observe that the integrands are 
expressions homogeneous of some degree in $u^\mu$.
Consider
\begin{equation}
I_{{\mu}_1\cdots{\mu}_p}
=
\frac{1}{\Omega_{d-2}}\int d\Omega_{d-2}\, u_{{\mu}_1}\cdots u_{{\mu}_p}\,.
\label
{solid-int-6D-2}
\end{equation}                                       
In the case of odd number of multiplying vectors $u^\mu$, the integrals vanish. 
If the number of multiplying vectors $u^\mu$ is even, then the 
integration are made through the use of 
the following formulas 
\begin{equation}
I_{{\mu}{\nu}}
=
-\left(\frac{1}{d-1}\right)
\stackrel{\scriptstyle v}{\bot}_{\hskip0.5mm\mu\nu}\,,
\label
{solid-it-6D-2}
\end{equation}                                       
\begin{equation}                                       
I_{\alpha \beta{\mu}{\nu}}
=
\frac{1}{
\left(d-1\right)\left(d+1\right)}\,\left(
\stackrel{\scriptstyle v}{\bot}_{\hskip0.5mm\mu\nu}\,
\stackrel{\scriptstyle v}{\bot}_{\hskip0.5mm\alpha\beta}+
\stackrel{\scriptstyle v}{\bot}_{\hskip0.5mm\alpha\mu}\,
\stackrel{\scriptstyle v}{\bot}_{\hskip0.5mm\beta\nu}+
\stackrel{\scriptstyle v}{\bot}_{\hskip0.5mm\alpha\nu}\,
\stackrel{\scriptstyle v}{\bot}_{\hskip0.5mm\beta\mu}\right),
\label
{solid-int-d-4}
\end{equation}                                       
\begin{eqnarray}
I_{\alpha \beta\gamma \lambda{\mu}{\nu}}
=
-\frac{1}{\left(d-1\right)\left(d+1\right)\left(d+3\right)}\,\biggl(
\stackrel{\scriptstyle v}{\bot}_{\hskip0.5mm\alpha\beta}\,
\stackrel{\scriptstyle v}{\bot}_{\hskip0.5mm\gamma\lambda}\,
\stackrel{\scriptstyle v}{\bot}_{\hskip0.5mm\mu\nu}
+
\stackrel{\scriptstyle v}{\bot}_{\hskip0.5mm\alpha\beta}\,
\stackrel{\scriptstyle v}{\bot}_{\hskip0.5mm\gamma\mu}\,
\stackrel{\scriptstyle v}{\bot}_{\hskip0.5mm\lambda\nu}
+
\stackrel{\scriptstyle v}{\bot}_{\hskip0.5mm\alpha\beta}\,
\stackrel{\scriptstyle v}{\bot}_{\hskip0.5mm\gamma\nu}\,
\stackrel{\scriptstyle v}{\bot}_{\hskip0.5mm\lambda\mu}
\nonumber\\
+
\stackrel{\scriptstyle v}{\bot}_{\hskip0.5mm\alpha\gamma}\,
\stackrel{\scriptstyle v}{\bot}_{\hskip0.5mm\beta\lambda}\,
\stackrel{\scriptstyle v}{\bot}_{\hskip0.5mm\mu\nu}
+
\stackrel{\scriptstyle v}{\bot}_{\hskip0.5mm\alpha\gamma}\,
\stackrel{\scriptstyle v}{\bot}_{\hskip0.5mm\beta\mu}\,
\stackrel{\scriptstyle v}{\bot}_{\hskip0.5mm\lambda\nu}\,
+
\stackrel{\scriptstyle v}{\bot}_{\hskip0.5mm\alpha\gamma}\,
\stackrel{\scriptstyle v}{\bot}_{\hskip0.5mm\beta\nu}\,
\stackrel{\scriptstyle v}{\bot}_{\hskip0.5mm\lambda\mu}\,
+
\stackrel{\scriptstyle v}{\bot}_{\hskip0.5mm\alpha\lambda}\,
\stackrel{\scriptstyle v}{\bot}_{\hskip0.5mm\beta\gamma}\,
\stackrel{\scriptstyle v}{\bot}_{\hskip0.5mm\mu\nu}
\nonumber\\
+
\stackrel{\scriptstyle v}{\bot}_{\hskip0.5mm\alpha\lambda}\,
\stackrel{\scriptstyle v}{\bot}_{\hskip0.5mm\beta\mu}\,
\stackrel{\scriptstyle v}{\bot}_{\hskip0.5mm\gamma\nu}
+
\stackrel{\scriptstyle v}{\bot}_{\hskip0.5mm\alpha\lambda}\,
\stackrel{\scriptstyle v}{\bot}_{\hskip0.5mm\beta\nu}\,
\stackrel{\scriptstyle v}{\bot}_{\hskip0.5mm\gamma\mu}
+
\stackrel{\scriptstyle v}{\bot}_{\hskip0.5mm\alpha\mu}\,
\stackrel{\scriptstyle v}{\bot}_{\hskip0.5mm\beta\nu}\,
\stackrel{\scriptstyle v}{\bot}_{\hskip0.5mm\gamma\lambda}
+
\stackrel{\scriptstyle v}{\bot}_{\hskip0.5mm\alpha\mu}\,
\stackrel{\scriptstyle v}{\bot}_{\hskip0.5mm\beta\gamma}\,
\stackrel{\scriptstyle v}{\bot}_{\hskip0.5mm\lambda\nu}
\nonumber\\
+
\stackrel{\scriptstyle v}{\bot}_{\hskip0.5mm\alpha\mu}\,
\stackrel{\scriptstyle v}{\bot}_{\hskip0.5mm\beta\lambda}\,
\stackrel{\scriptstyle v}{\bot}_{\hskip0.5mm\gamma\nu}
+
\stackrel{\scriptstyle v}{\bot}_{\hskip0.5mm\alpha\nu}\,
\stackrel{\scriptstyle v}{\bot}_{\hskip0.5mm\beta\mu}\,
\stackrel{\scriptstyle v}{\bot}_{\hskip0.5mm\gamma\lambda}
+
\stackrel{\scriptstyle v}{\bot}_{\hskip0.5mm\alpha\nu}\,
\stackrel{\scriptstyle v}{\bot}_{\hskip0.5mm\beta\lambda}\,
\stackrel{\scriptstyle v}{\bot}_{\hskip0.5mm\gamma\mu}\,
+
\stackrel{\scriptstyle v}{\bot}_{\hskip0.5mm\alpha\nu}\,
\stackrel{\scriptstyle v}{\bot}_{\hskip0.5mm\beta\gamma}\,
\stackrel{\scriptstyle v}{\bot}_{\hskip0.5mm\lambda\mu}
\biggr),
\label
{solid-int-d-6}
\end{eqnarray}
which are readily derived (see, e.~g., \cite{k06}).  
Here, 
\begin{equation}
\stackrel{\scriptstyle v}{\bot}_{\hskip0.5mm\mu\nu}
=
\eta_{\mu\nu}
-v_\mu v_\nu
\label
{projector-bot}
\end{equation}                                       
 is the operator that projects vectors
onto a hyperplane with normal $v^\mu$,
The number of terms in such decompositions of 
 $I_{{\mu}_1\cdots{\mu}_k}$
proliferates with $k$:
$I_{{\mu}_1\cdots{\mu}_4}$ 
contains $3$ monomials 
$\stackrel{\scriptstyle v}{\bot}\,\stackrel{\scriptstyle v}{\bot}$,
$I_{{\mu}_1\cdots{\mu}_6}$ 
involves $3\cdot 5$ monomials 
$\stackrel{\scriptstyle v}{\bot}\,\stackrel{\scriptstyle v}{\bot}\,
\stackrel{\scriptstyle v}{\bot}$,
$I_{{\mu}_1\cdots{\mu}_8}$ 
comprises $3\cdot 5\cdot 7$ monomials 
$\stackrel{\scriptstyle v}{\bot}\,\stackrel{\scriptstyle v}{\bot}\,
\stackrel{\scriptstyle v}{\bot}\,\stackrel{\scriptstyle v}{\bot}$, etc. 
If $k\ge 6$, then calculations with $I_{{\mu}_1\cdots{\mu}_k}$ are rather 
tedious, so that we restrict our discussion to the 
dimensions $d=4$ and $d=6$.
In these cases we need only handling $I_{{\mu}{\nu}}$ and $I_{\alpha\beta{\mu}{\nu}}$.

Using the identities
\begin{equation}
v^2=1,
\quad
\left(v\cdot a\right)=0,
\quad
\left(v\cdot {\dot a}\right)=-a^2,
\label
{identities}
\end{equation}                       
we find from (\ref{mf-6}) and (\ref{mfr-8}) that
\begin{equation}
\left({\mathfrak b}^{(6)}\right)^2
=
\left({a}\cdot u\right)^2+
{a^2},
\label
{mf-6-sqr}
\end{equation}                       
\begin{equation}
\left({\mathfrak b}^{(8)}\right)^2
=
\left[(\stackrel{\scriptstyle v}
{\bot}{\dot a})^2+9\,(a\cdot u)^2a^2+9\,(a\cdot u)^4+({\dot a}\cdot u)^2\right]
-
3\left[(a^2)^.\,(a\cdot u) +2\,(a\cdot u)^2({\dot a}\cdot u)\right].
\label
{tube-int-6D}
\end{equation}                                       
Thus, in four and six dimensions,
the radiation rate is given, respectively, by
\begin{equation}
{\dot{\cal P}}_{\mu}^{(4)}
=
-\frac{2}{3}\,
a^{2}
v_\mu 
\label
{radiat-rate-4}
\end{equation}                                       
and
\begin{equation}
{\dot{\cal P}}_{\mu}^{(6)}
=
\frac{1}{9}\,\frac{1}{5\cdot 7}\left\{{4}\left[{
{16}\,(a^2)^2 -7\,\dot a}^2\right]v_\mu-{3}\cdot{5}\left(a^2\right)^.\,a_\mu 
+{6}\,a^2
(\stackrel{\scriptstyle v}{\bot}{\dot a})_\mu\right\}.
\label
{P-rad-6D}
\end{equation}

\section{Discussion and outlook}
\label{concl}
Let us summarize our discussion of the methods for obtaining 
retarded field configurations due to 
a single point charge in
$2n$-dimensional Minkowski spacetime.
The retarded Green's function technique
is presently accepted as
the standard approach.
Iwanenko and Sokolow \cite{IwanenkoSokolow}
pioneered the use of this technique.
The approach based on the ans{\"a}tze of a particular form,
such as those defined in (\ref{2-Ansatz}), (\ref{4-Ansatz}), 
and (\ref{6-Ansatz}), was developed in Ref.~\cite{k99}.
This procedure for solving Maxwell's equations (without resort to 
Green's functions) is found to be of particular assistance in solving the 
Yang--Mills equations \cite{k06}.
It seems likely that the tool of greatest practical utility involves 
the notion of prepotential, in particular the simplest way for 
calculating the retarded vector 
potential ${\cal A}^{(2n)}_{\mu}$ is given by Eq.~(\ref{A-2p+2-A-2-diff}). 

Close inspection of exact solutions to $d$-dimensional
Maxwell's equations shows that 
the retarded field 
strength ${\cal F}^{(2n)}_{\mu\nu}$ 
generated by a point charge living in a
$2n$-dimensional 
world
is expressed in terms of the 
retarded vector potentials ${\cal A}^{(2m)}_{\mu}$ 
due to this charge   
in
$2m$-dimensional 
worlds nearby, Eqs.~(\ref{F-2})--(\ref{F-10}).
The fact that the state of the 
retarded electromagnetic field in a given even-dimensional manifold is 
entangled with those of contiguous even-dimensional manifolds may be the 
subject of far-reaching philosophical speculations.
To illustrate, it follows from (\ref{F-4}) that, while living in $d=4$, a
charge feels
a specific impact from
$d=2$ and $d=6$. 
The responsibility for this entanglement may rest with either coexistence 
on an equal footing of different $2p$-branes in some braneworld scenario 
or manifestation of contiguous 
`parallel' realms.

A notable feature of Eqs.~(\ref{F-2})--(\ref{F-10}) is that the world line 
$z^\mu(s)$ of the charge 
generating these field configurations is described by different numbers of the
principal curvatures $\kappa_j$ for different spacetime dimensions.
To be specific, we refer to Eq.~(\ref{F-4}).
The world line appearing in ${\cal A}^{(2)}_{\mu}$ is
a planar curve, specified solely by $\kappa_1$, while
that appearing in ${\cal A}^{(6)}_{\mu}$ is
a curve characterized (locally) by five essential parameters $\kappa_1$,
 $\kappa_2$, $\kappa_3$,  $\kappa_4$, $\kappa_5$.  
If we regard the world line $z^\mu(s)$ in ${\mathbb M}_{1,2n-1}$ as the basic
object, then both projections of this curve onto lower-dimensional spacetimes
and its extensions to higher-dimensional spacetimes are rather arbitrary.
Nevertheless,  Eqs.~(\ref{F-2})--(\ref{F-10}) are invariant under 
variations of these mappings of the world line $z^\mu(s)$.

The advanced fields ${\cal F}_{\rm adv}$ 
can be also represented as the sums of exterior products of $1$-forms
${\cal A}_{\rm adv}$ similar 
to (\ref{F-2})--(\ref{F-10}), 
whereas combinations $\alpha\,{\cal F}_{\rm ret}+\beta\,{\cal F}_{\rm adv}$,
$\alpha\beta\ne 0$, are not.
Therefore, Eqs. (\ref{F-2})--(\ref{F-10}) do not hold for 
field configurations 
satisfying the St{\"u}ckelberg--Feynman boundary condition.
We thus see that the remarkably simple
structures displayed in Eqs.~(\ref{F-2})--(\ref{F-10}) are inherently classical.

Based on Eqs.~(\ref{F-2})--(\ref{F-10}), we put the rate of radiated 
energy-momentum of electromagnetic field in a compact form, 
Eqs. (\ref{radiat-rate}) and (\ref{rad-rate}). 
Let us recall that there are two alternative concepts of radiation, proposed by
Dirac and Teitelboim (for a review see \cite{k92});
the latter was entertained in Sec.~3.  
Although these concepts have some points in common, they are not equivalent.
Accordingly, the fact [afforded by (\ref{energy-tensr-rad}) and 
(\ref{radiat-rate})] that the
radiation in  $2n$-dimensional spacetime is an 
infrared phenomenon stemming from the next even dimension $d=2n+2$
cannot be clearly recognized until the Teitelboim's definition of radiation 
is invoked.

Why is it essential to draw the stress-energy tensor for 
introducing the concept of radiation?
It is still common to see the assertion that the degrees of freedom 
related to the
radiation may be identified directly in ${\cal F}^{(2n)}$ if one takes the piece 
of ${\cal F}^{(2n)}$ shown in (\ref{infrared-F-}) as the
`radiation field'.
However, this assertion is erroneous.
First, the construction ${{\cal A}}^{(2)}\wedge{\bar{\cal A}}^{(2n+2)}$ which
allegedly plays the role of radiation field
is in no sense dynamically 
independent of the rest of ${\cal F}^{(2n)}$.
Second, the bivector  $\varpi={{\cal A}}^{(2)}\wedge{{\cal A}}^{(2n+2)}$ 
is
deprived of information about the vector ${\bar{\cal A}}^{(2n+2)}_\mu$.
A pictorial view of $\varpi$
is the  parallelogram of the vectors ${{\cal A}}^{(2)}_\mu$ and
 ${{\cal A}}^{(2n+2)}_\mu$.
The bivector  $\varpi$  
is 
independent 
of concrete directions and magnitudes of the constituent vectors 
${{\cal A}}^{(2)}_\mu$ and
 ${{\cal A}}^{(2n+2)}_\mu$; $\varpi$ depends only on the parallelogram's 
orientation
and area  
$
{\mathfrak S}
=
 |{{\cal A}}^{(2)}\cdot{{\cal A}}^{(2n+2)}|$.
By virtue of (\ref{constr}), ${\bar{\cal A}}^{(2n+2)}_\mu$ makes no
contribution to ${\mathfrak S}$. 
It can be shown (much as was done in \cite{k06}, p.~181) that the 
term  scaling as $\rho^{1-n}$ can be
eliminated by a local SL$(2,{\mathbb R})$ transformation of the plane spanned
by the vectors
${{\cal A}}^{(2)}$ and ${{\cal A}}^{(2n+2)}$
which leaves 
the bivector  $\varpi$ invariant.
In other words, there is a reference frame in which
the `radiation field'  (\ref{infrared-F-}) vanishes over all spacetime
(except for the future null infinity).

The implication of this argument is that the radiation is determined not 
only by the retarded field ${\cal F}^{(2n)}$
as such but 
also by the frame of reference in which ${\cal F}^{(2n)}$ is measured.  
On the other hand,  the stress-energy tensor $\Theta^{\mu\nu}$ 
is not invariant under such SL$(2,{\mathbb R})$ transformations.
 $\Theta^{\mu\nu}$ carries information about both
the field ${\cal F}^{(2n)}$ and the frame which is used to 
describe ${\cal F}^{(2n)}$.



\begin{thebibliography}{99}

\bibitem
{Arkani-Hamedb}
N. Arkani-Hamed, S. Dimopoulos, and G. Dvali,
``The Hierarchy problem and new dimensions at a
millimeter,''
{\it Phys. Lett.}
{\bf B 429}  (1998) 263;
hep-th/9803315.


\bibitem
{Antoniadis}
I. Antoniadis, N. Arkani-Hamed, S. Dimopoulos, and G. Dvali,
``New dimensions at a millimeter to a Fermi
and superstrings at a TeV,''    
{\it Phys. Lett.}
{\bf B 436}  (1998) 257;
hep-th/9804398.

\bibitem
{Arkani-Hamedb-prd}
N. Arkani-Hamed, S. Dimopoulos, and G. Dvali,
``Phenomenology, astrophysics and
cosmology of theories with sub-millimeter dimensions and TeV scale quantum
gravity,''
{\it Phys. Rev.}
{\bf D 59} (1999) 086004; hep-th/9804398.

\bibitem
{Akama}
K. Akama, {``Pregeometry,''} in {\it Gauge Theory 
and Gravitation},
 Proceedings, Nara, 1982, edited by K. Kikkawa, N. Nakanishi 
and H. Nariai, Lecture Notes in Physics, {\bf 176}, 
 (Springer, Berlin, 1983) pp. 267-271; hep-th/0001113.


\bibitem
{Rubakov}
V. A. Rubakov and M. E. Shaposhnikov, 
``Do we live inside a domain wall?,''
{\it Phys. Lett.}
{\bf B 125}  (1983)  136.

\bibitem
{Visser}
M. Visser, 
``An exotic class of Kaluza-Klein models,'' 
{\it Phys. Lett.}
{\bf B  159}  (1985) 22.

\bibitem
{Randall}
L. Randall and R. Sundrum, 
``A large mass hierarchy from a small extra dimension,''
{\it Phys. Rev. Lett.}
{\bf  83} (1999) 3370; hep-ph/9905221.

\bibitem
{Randall-s}
L. Randall and R. Sundrum, 
``An alternative to compactification,'' 
{\it Phys. Rev. Lett}.
{\bf 83} (1999) 4690; hep-th/9906064.

\bibitem
{Rubakov-usp}
 V. A. Rubakov, ``Large and infinite extra dimensions: An introduction,'' 
{\it Phys. Uspekhi}.
{\bf 44}  (2001) 871; hep-ph/0104152.

\bibitem
{Maartens}
R. Maartens, 
``Brane world gravity,'' 
{\it Living Rev. Rel}. {\bf 7} (2004) 7; gr-qc/0312059.

\bibitem
{k92}
B.\ P.\ Kosyakov, 
``Radiation in electrodynamics and  the Yang--Mills theory,'' 
{\it Sov. 
Phys.---Uspekhi},
{\bf 35} (1992) 135.
 
\bibitem
{k06}  
B. Kosyakov,
{\it Introduction to the Classical Theory of Particles and Fields}
(Springer, Berlin, 2007).

\bibitem
{Mironov-Morozov}
A.~Mironov and A.~Morozov, 
``On the problem of radiation friction beyond 4 and 6 dimensions,''
 hep-th/0710.5676v1.


\bibitem
{k99}
B. P. Kosyakov, 
``Exact solutions of classical electrodynamics and the Yang--Mills--Wong theory
in even-dimensional spacetime,''
{\it Theor. Math. Phys}. {\bf 119} (1999) 493;  hep-th/0207217.

\bibitem
{Gal'tsov}  
D.~{Gal'tsov},
``Radiation reaction in various dimensions,''  
{\it Phys. Rev.}
{\bf D~66}  (2002) 025016; hep-th/0112110.

\bibitem
{Kazinski}  
P.~{Kazinski}, S.~Lyakhovich, and A.~Sharapov,  
``Radiation reaction and renormalization in classical electrodynamics of a point particle in any 
dimension,''  
{\it Phys. Rev.}
{\bf D~66}  (2002) 025017; hep-th/0201046.

\bibitem
{Yaremko}  
Yu.~{Yaremko},  
``Radiation reaction, renormalization and conservation laws in six-dimensional classical electrodynamics,''  
{\it J. Phys.}
{\bf A~37}  (2004) 1079.

\bibitem
{Sarioglu}  
M. G{\"u}rses and {\"O}. Sario{\u g}lu, 
``Li\'enard--Wiechert potentials in even dimensions,'' 
{\it J.\ Math.\ Phys}.\ {\bf 44} (2003) 4672;  hep-th/0303078v2.

\bibitem
{MironovMorozov}
A.~Mironov and A.~Morozov, 
``Radiation beyond four-dimension space-time,''
  hep-th/0703097v1.

\bibitem
{IwanenkoSokolow}  
D. Iwanenko {\&} A.\ Sokolow,
{\it Die Klassische Feldtheorie} 
(Akademie, Berlin, 1953).
Translated from the 
Russian edition 1951. 

\bibitem
{Teitelboim}
C.\ Teitelboim,
``Splitting of Maxwell tensor: Radiation reaction without advanced 
fields,'' 
{\it Phys. Rev.}
{\bf D 1}  (1970) 1572.


\end{thebibliography}
\end{document}